\documentclass[conference]{IEEEtran}
\IEEEoverridecommandlockouts
\usepackage{cite}
\usepackage{amsmath,amssymb,amsfonts}
\usepackage{algorithmic}
\usepackage{graphicx}
\usepackage{amsmath}
\usepackage{amsfonts}
\usepackage{textcomp}
\usepackage{siunitx} 
\usepackage{xcolor}
\def\BibTeX{{\rm B\kern-.05em{\sc i\kern-.025em b}\kern-.08em
    T\kern-.1667em\lower.7ex\hbox{E}\kern-.125emX}}
\begin{document}

\title{Automating Capacitor Part Selection with Dual-Objective Optimization\\}
\author{\IEEEauthorblockN{Luke Brantingham}
\IEEEauthorblockA{\textit{Google}\\
Chicago, IL \\
lbrantingham@google.com}
\and
\IEEEauthorblockN{Jason Grover}
\IEEEauthorblockA{\textit{Google}\\
Chicago, IL \\
jgrover@google.com}
}

\maketitle

\begin{abstract}
This paper presents a novel framework for optimizing capacitor selection in electronic design using multi-objective linear constrained optimization techniques. We demonstrate the effectiveness of this approach in minimizing cost and board area while meeting critical performance requirements and extend the framework to an economic model of optimal capacitor utilization at the design- or multi-design level.
\end{abstract}

\section{Introduction}

Multi-layer ceramic capacitors (MLCCs) are essential components in modern electronics, serving critical functions in power, RF, and analog circuits. The rapidly expanding MLCC market \cite{mlccmarket} offers designers a vast selection of package sizes, voltage ratings, and performance characteristics. However, effectively navigating this complex design space to optimize component selection presents a significant challenge.

Capacitor selection often involves balancing competing objectives: minimizing cost and minimizing board area. These objectives are typically in tension, and traditional manual selection methods may not identify optimal solutions.

While the selection methods in this paper do not completely replace human intervention in all cases, several simple capacitor selection tasks are found to be solvable or provide the human designer an efficient starting-point. The scope of the models presented in this paper include optimization of cost and placement area for the following design settings:

\begin{itemize}
  \item $C_{\text{eff}}$ (a minimum derated capacitance requirement),
  \item $|Z_i|$ targets (impedance envelope),
  \item Combined PDN (power distribution network) model with $|Z_i|$ targets.
\end{itemize}

We believe that certain design contexts benefit from the use of each of these formulations. However, the scope of this article is not to rigorously document the potential applications.

\section{Selecting MLCCs for $C_{\text{eff}}$ requirements}

\subsection{Defining the capacitance constraint}

When designing power systems, it's common to encounter specifications that require a minimum effective capacitance $C_{\text{eff}}$ , as seen in datasheets like \cite{tidatasheet} and \cite{adidatasheet}. This value represents the capacitance of a component after accounting for derating, which is a reduction in capacitance due to factors like DC voltage bias. To ensure designs meet these requirements, engineers often keep a database of $C_{\text{eff}}$  values for various capacitors across different DC bias levels. This allows them to select a combination of components that meet the necessary capacitance requirement.

The task is to select some mix of capacitors that together have at least $C_{\text{eff}}$ total derated capacitance. For $I$ possible capacitor part types, this constraint can be written as:

\begin{equation}\label{capconstraint}
\begin{split}
    \sum_{i}^{I} C_iN_i\geq C_{\text{eff}} \\
    \text{where}\quad N_i \in \mathbb{Z} \quad \forall i \\
      \text{and} \quad N_i \geq 0  \quad  \forall i
\end{split}
\end{equation}

where $C_i$ is the derated capacitance of part $i$ at the DC bias voltage target and $N_i$ is the number of part $i$ in the solution set.

The available parts $i$ should be pre-filtered by application based on usability in that setting. Common design filters of this type include:

\begin{itemize}
  \item Maximum part height to avoid interference with other parts in the product,
  \item Minimum part voltage rating selected to maximize the useful life of the part \cite{murata} \cite{nasa},
  \item Temperature, voltage and aging stability of capacitance (influencing dielectric material choice),
  \item Approved manufacturers in the supply chain setting.
\end{itemize}

\subsection{Defining our preference model}

\begin{figure}[htbp]
\centerline{\includegraphics{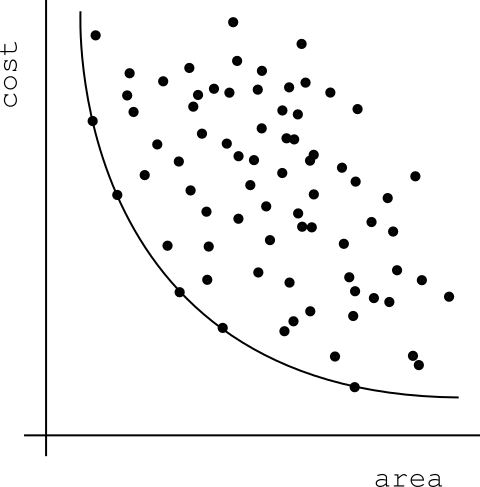}}
\caption{Simplified expected solution space for capacitor optimization problems with Pareto frontier drawn.}
\label{figconvex}
\end{figure}

Beyond the electrical constraints, electronics manufacturers pursue dual minimization objectives: cost and board area of the solution. As volumetric density of capacitance generally increases capacitor cost, we expect the efficient frontier of the feasible region of our optimization to be convex, as shown in Figure \ref{figconvex}.

Each dot represents a unique solution to the capacitance constraint \eqref{capconstraint}. Along the Pareto frontier shown, different solutions might be chosen that match the desired trade-offs and goals of the design. For example, a smart watch design might choose a smaller-area design at higher cost than would a server rack design where space is not at a premium.

One approach to formalizing the different preferences applications might have is a simple linear scalarization, where we minimize the objective function:

\begin{equation}\label{objfn}
    \sum_{i}^{I} (Ka_i+b_i)N_i
\end{equation}

where $a_i$ represents the economic cost - in currency - of capacitor part $i$ and $b_i$ represents the capacitor placement area. The value of $K$ represents the designer's willingness to trade off cost with solution area. It is a fixed value that the designer provides the model. Smaller values of $K$ should be used in situations where designers are more willing to pay for miniaturization.

Beyond cost and area as objectives for capacitor selection, our objective function could also include weighted operational dimensions which out of the scope of this paper, such as :

\begin{itemize}
  \item Part sourcing risk,
  \item Footprint compatibility with possible substitutes (modeled as an option-to-switch value),
  \item A factor adjusting for the discount of commonizing part selection across applications within the same product.
\end{itemize}

\subsection{Solving with examples}

Using standard integer-constrained linear optimization methods, we can easily minimize \eqref{objfn} under the constraint of \eqref{capconstraint}. Our decision variables are each $N_i$, the number of each type of capacitor part in the solution. 

As a thematic example, we will satisfy $C_{\text{eff}} \geq 4\mu F$ with the following options derated at 3.3V:

\begin{table}[h!]
\caption{Hypothetical capacitor optimization options}
\begin{center}
\begin{tabular}{|c|c|c|c|c|}
\hline
\textbf{part \#} & \textbf{Description} & \textbf{$\mu F$ at 3.3V} & \textbf{Cost} & \textbf{Area mm$^2$} \\
\hline
A       & 1$\mu F$ 0201 6.3V   & 0.35                                             & \$0.002                           & 0.7                                                  \\
\hline
B       & 2.2$\mu F$ 0201 6.3V & 0.85                                             & \$0.003                           & 0.7                                                  \\
\hline
C       & 1$\mu F$ 0201 10V    & 0.45                                             & \$0.003                           & 0.7                                                  \\
\hline
D       & 2.2$\mu F$ 0201 10V  & 0.95                                             & \$0.004                           & 0.7                                                  \\
\hline
E       & 2.2$\mu F$ 0402 6.3V & 0.90                                              & \$0.003                           & 1.3                                                  \\
\hline
F       & 4.7$\mu F$ 0402 6.3V & 1.70                                              & \$0.007                           & 1.3                                                  \\
\hline
G       & 2.2$\mu F$ 0402 10V  & 1.00                                                & \$0.005                           & 1.3                                                  \\
\hline
H       & 4.7$\mu F$ 0402 10V  & 1.95                                             & \$0.008                           & 1.3                                                 \\
\hline
\end{tabular}
\label{tab1}
\end{center}
\end{table}

Subject to the objective function \eqref{objfn}, we can sweep $K$ to simulate a variety of designer preferences, and calculate an ideal capacitor strategy with a branch-and-bound linear integer optimization solver. The result of this optimization step is seen in table \ref{tab:opresults}.

\begin{figure*}[h]
\centering
\setlength\fboxsep{0pt}
\setlength\fboxrule{0pt}
\fbox{\includegraphics[width=6.2in]{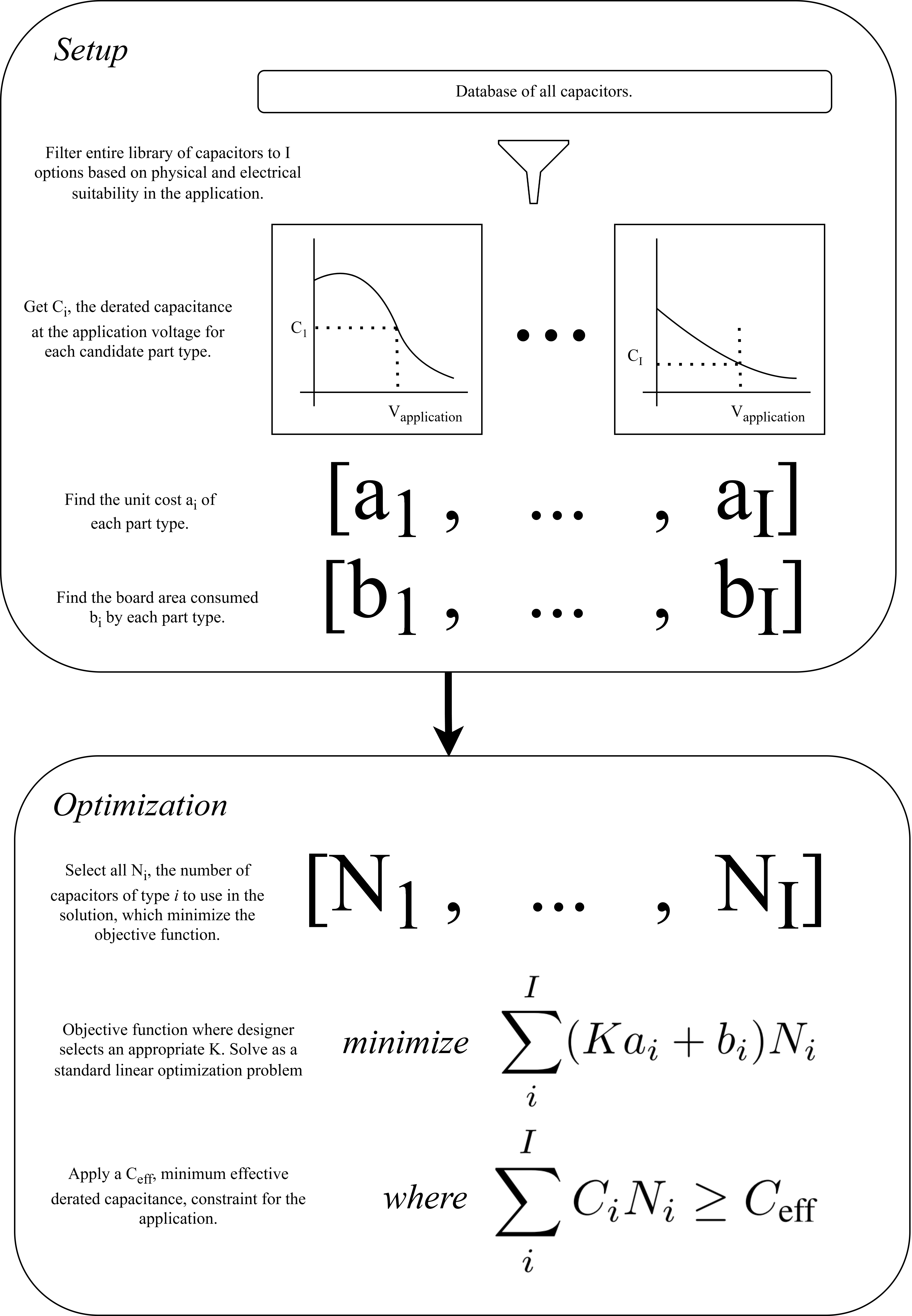}}
\caption{The optimization process for $C_{\text{eff}}$ optimization.}
\label{fig:process}
\end{figure*}

For a graphical representation of this optimization process, see figure \ref{fig:process} at the end of this paper.

\begin{table}[htbp]
\caption{Example capacitor optimization solutions based on preference}
\begin{center}
\begin{tabular}{|c|c|c|c|c|}
\hline
\textbf{K}  & \textbf{0.5}   & \textbf{1}  & \textbf{2}                                                  \\
\hline
$N_A$  & 0   & 0 & 0                                               \\
\hline
$N_B$  & 1   & 3 & 5                                                  \\
\hline
$N_C$  & 0   & 0 & 0                                                  \\
\hline
$N_D$  & 0   & 0 & 0                                                  \\
\hline
$N_E$  & 0   & 0 & 0                                                \\
\hline
$N_F$  & 2   & 1 & 0                                                \\
\hline
$N_G$  & 0   & 0 & 0                                                 \\
\hline
$N_H$  & 0   & 0 & 0                                                  \\
\hline
\end{tabular}
\label{tab1}
\end{center}
\label{tab:opresults}
\end{table}

\section{Selecting capacitors for impedance envelopes}

\subsection{Defining the impedance envelope problem}

When capacitors are used in filtering applications across wide a frequency spectrum, the \textit{impedance envelope} model is often used for selecting capacitors and simulating power delivery. 

The impedance envelope is a function $Z(f)$ that represents the minimum impedance to ground a port requires. The impedance includes paths through copper, regulator control loops, and passives. Capacitor selection addresses the mid-frequencies of the PDN - between the regulator and PCB layout regions. Impedance envelopes are often derived from voltage range specifications ($V_{min}$, $V_{max}$) at power rail loads \cite{tiguide}.

Whether selecting capacitors manually or with computer-aided optimization methods, it is a prerequisite task of the designer to isolate the targeted impedance mask region to address with capacitors. Once this is done, the designer can represent the capacitor-region mask into $M$ discrete terms:

\begin{equation}\label{impedanceconstraint}
\begin{split}
    |Z|_{f_m} < T_{Z@f_m} \\
    \text{for }m=1,2,\ldots,M 
\end{split}    
\end{equation}

The impedance magnitude of the capacitor solution at frequency $f_{n}$ must be lower than the target impedance mask value $T_{Z@f_m}$ for all $n$ discrete impedance mask points.

Ignoring the real effects of resonance, as is typically done in the manual capacitor selection phase of design, we can leverage the following relationship between parallel capacitors:

\begin{equation} \label{parallelapprox}
\begin{split}
    \frac{1}{|Z|_{f_m}} \approx \sum_{i}^{I} \frac{N_i}{|Z|_{i@f_m}} \\
    \text{for }m=1,2,\ldots,M
\end{split}
\end{equation}

The approximation \eqref{parallelapprox} combines the parallel impedances at $f_m$ of the solution set into a single effective impedance at $f_m$. Although this sets us up to solve the same kind of optimization problem as in section I, notice that \eqref{parallelapprox} is non-linear, so it cannot be used directly.

Instead, we will use \textit{admittance} - the reciprocal of impedance - in order to solve the system with fast linear methods:

\begin{equation} \label{admittanceconstraint}
\begin{split}
    |Y|_{f_m} > T_{Y@f_m}, \\
    \text{for }m=1,2,\ldots,M 
\end{split}
\end{equation}

\begin{equation}  \label{admittancesum}
\begin{split}
    |Y|_{f_m} \approx \sum_{i}^{I} N_i|Y|_{i@f_m}\\
    \text{for }m=1,2,\ldots,M 
\end{split}
\end{equation}

With \eqref{admittanceconstraint} and \eqref{admittancesum}, we add $M$ more constraint inequalities to our linear system, but we can solve it identically, optimizing for a weighted scalar objective function trading off cost and area.

\subsection{Summarizing a mixed impedance envelope and $C_{\text{eff}}$ linear model}

A common challenge faced by electronic designers involves selection of capacitors on a power rail subject to the minimum derated capacitance requirement of a voltage regulator and an impedance envelope demanded by the power rail's load(s).

\begin{figure}[htbp]
\centerline{\includegraphics[width=8cm]{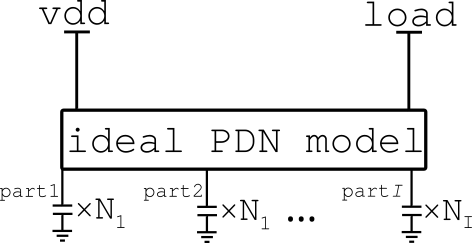}}
\caption{Regulator-and-load topology with ideal PDN.}
\label{figidealcircuit}
\end{figure}

We will assume an ideal PDN model, that is, no parasitic impedance from $vdd$ to $load$, or from capacitor to capacitor.

Our optimization problem is formulated as:

\begin{equation} \label{linearsummary}
\begin{split}
    \text{select $N_i$ for $i=1,2,\ldots,I$} \\
    \text{in order to minimize }\sum_{i}^{I} (Ka_i+b_i)N_i, \\
    \text{subject to } \\
    \sum_{i}^{I} C_iN_i\geq C_{\text{eff}} \text{ and} \\
    \sum_{i}^{I} N_i|Y|_{i@f_m} > T_{Y@f_m} \text{  for } m=1,2,\ldots,M 
\end{split}
\end{equation}

The simple, idealized model of \eqref{linearsummary} will produce results for $N_i$ which can inform the electronic designer's capacitor selection strategy.

\subsection{Mixed impedance envelope capacitor selection example}

As a stylistic example, we will consider a capacitor selection challenge with the following constraints:

\begin{equation} \label{stylistic}
\begin{split}
    \text{Select $N_i$ for $i=1,2,\ldots,I$} \\
    \text{in order to minimize }\sum_{i}^{I} (Ka_i+b_i)N_i, \\
    \text{subject to the constraints:} \\
    \sum_{i}^{I} C_iN_i\geq \text{12 $\mu$F at V=1.15 Volts,} \\
    \sum_{i}^{I} N_i|Y|_{i\text{@100KHz}} > \frac{1}{0.1 \Omega}, \\
    \sum_{i}^{I} N_i|Y|_{i\text{@1MHz}} > \frac{1}{0.01 \Omega}, \\
    \sum_{i}^{I} N_i|Y|_{i\text{@10MHz}} > \frac{1}{0.005 \Omega}, \\
    \sum_{i}^{I} N_i|Y|_{i\text{@100Mhz}} > \frac{1}{0.01 \Omega}, \\
    \sum_{i}^{I} N_i|Y|_{i\text{@1GHz}} > \frac{1}{0.1 \Omega} \\
\end{split}
\end{equation}

With a library of $I=$ several hundred available parts, we can solve this linear optimization problem for a given $K$ in under 10 milliseconds with standard linear solvers.

By sweeping $K$ in 40 log-spaced steps between 0.01 and 100, we can generate a number of efficient solutions which can be plotted in cost, area 2D space as an efficient frontier:

\begin{figure}[htbp]
\centerline{\includegraphics[width=8cm]{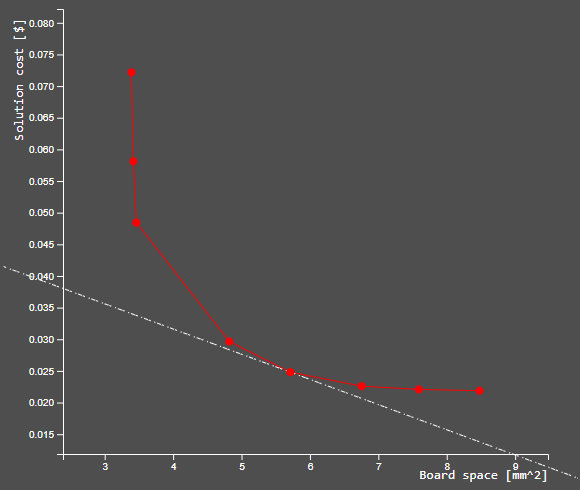}}
\caption{Efficient frontier solutions found for the optimization problem \ref{stylistic} using a real reference part library. The tangency line for $K=2.51$ (the designer is willing to save a penny by increasing solution size by 2.51 square millimeters).}
\label{figefficient}
\end{figure}

Each of the red dots in Figure \ref{figefficient} is a unique solution $N_i$ for $i=1,2,\ldots,I$. The solutions shown range from having 2 to 5 unique capacitor parts included ($N_i > 0$ for 2 to 5 unique $i$), and up to 14 total parts ($\sum_{i}^{I} N_i >= 14$). From our experience, most generated solutions are unlikely to be found by human search.

Since the solutions are generated so quickly, designers can easily run ad-hoc studies analyzing the cost (in cents and board space) to a solution by doing the following and re-running the optimization:

\begin{itemize}
  \item adding a new part to the database,
  \item restricting the maximum height of the solution,
  \item modifying the preference between cost and board space ($K$).
\end{itemize}

While this process greatly automates the search for solutions, each solution must be validated with real resonance and PDN modelling.

\subsection{Adding a simple non-ideality to the PDN model}

Our model in the above section will optimize the selection set within an unrealistic assumption of an ideal PDN. The physical layout of the product's conductors will define the impedance of our PDN. Given a mature design, we can simulate all of the relevant port-to-port PDN impedances with numerical methods, as is standard in the industry. 

Early in the product development stage, we may choose to implement a crude approximation of our PDN. In modern electronics, capacitors are often placed on the side of the PCB opposite the load port, making the capacitor placement-to-load impedance dominated by the series impedance of the via structure between them.

\begin{figure}[htbp]
\centerline{\includegraphics[width=8cm]{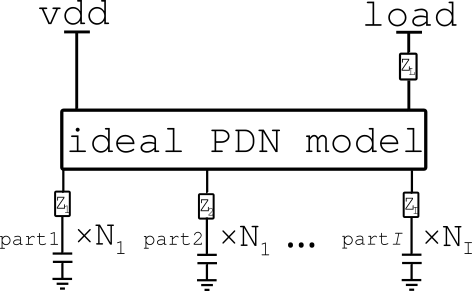}}
\caption{Regulator-and-load topology with capacitor and load series impedance.}
\label{figidealcircuit}
\end{figure}

In the above schematic, we will use the $Z_m$ values to pre-transform the individual part's effective impedances:

\begin{equation} \label{ztransform}
\begin{split}
|Z|_{i@f_m}^* = |Z|_{i@f_m} + Z_m\\
    \text{for }m=1,2,\ldots,M 
\end{split}
\end{equation}

We will also pre-transform the load impedance targets: 

\begin{equation}\label{impedanceconstraint}
\begin{split}
    T_{Z@f_m}^* =  T_{Z@f_m}^* - Z_L\\
    \text{for }m=1,2,\ldots,M 
\end{split}    
\end{equation}

We observe that if $Z_L > T_{Z@f_m}$, the solution becomes infeasible, as expected. Re-solving the linear optimization problem with $T_{Z@f_m}^*$ and $|Z|_{i@f_m}^*$, we achieve a capacitor selection set that is constrained by a more realistic set of PDN targets without introducing any non-linearity into the model.

\section{Selecting capacitors within more robust PDN models}

\subsection{Basic placement location-informed optimization models}

Further elaborating on our ideal PDN model, we can model a parasitic impedance $Z_{jk}$ between candidate placement locations $j$ and $k$. Having introduced placement locations into our model, it is useful to also specify impedance spec requirements per location, as would be the case for a regulator with multiple loads connected by a shared PDN. Each capacitor placement area affects the effective impedance at every other location's load as well as its own. 

\begin{figure}[htbp]
\centerline{\includegraphics[width=8cm]{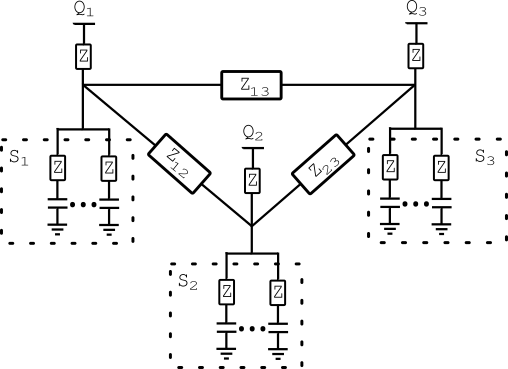}}
\caption{Placement-location selection problem schematic.}
\label{figidealcircuit}
\end{figure}

In general, we can think of $J$ possible locations for capacitor placement. Each spec location $Q_j$ in the diagram has its own impedance envelope requirement:

\begin{equation}\label{impedanceconstraint}
\begin{split}
    |Z|_{f_m@j} < T_{Z@f_m@j} \\
    \text{for }m=1,2,\ldots,M \\
    \text{for }j=1,2,\ldots,J 
\end{split}    
\end{equation}

Opportunities for capacitor placement with no corresponding local impedance mask requirement can be modeled with an infinite impedance mask. This tactic can be used when considering placement of capacitors in a region of the PCB central to several loads or distant from any load in particular.

Our number of decision variables gets multiplied by the number of candidate placement areas. We now have $I \times J$ decision variables, each notated $N_{ij}$.

In product design, though each capacitor costs the same (economically speaking) per location, we may value the placement area more dearly in one location than the other. For this reason, we can model our objective to minimize function as 

\begin{equation}\label{pdnobjfn}
    \sum_{j}^{J}\sum_{i}^{I} (K_ja_i+b_i)N_{ij}
\end{equation}

where the designer assigns the values of $K_1, K_2, ... K_J$ according to the relative cost of placement area in that location. For instance, a relatively congested placement area $j$ will have a relatively small $K_j$.

For the following modelling, we will assume the pre-transformations of section III.D have already been made. By solving the impedance circuit, we can model our impedance envelope constraints as:

\begin{equation}\label{pdnimpedanceconstraint}
\begin{split}
    |Y|_{f_m@j} > T_{Y@f_m@j} \\
    \text{for }m=1,2,\ldots,M \\
    \text{for }j=1,2,\ldots,J \\
    \text{where } |Y|_{f_m@j} = \sum_{i}^{I}N_{ij}Y_{i@f_m} + \sum_{k \neq j}^{J}\frac{1}{\frac{1}{Y_{jk}} + \frac{1}{\sum_{i}^{I}N_{ik}Y_{i@f_m}}}
\end{split}    
\end{equation}

This does not seem to us to be linearizable, either in impedance or admittance forms. We are left with a mixed-integer non-linear (smooth) programming task (MINLP).

Solving this optimization problem with available open source solvers such as gekko \cite{gekko} yields promising solutions that automatically trade off designer placement preferences with cost and area of the solution.

\subsection{Expanding the design search space to include PDN impedances}

Equation \eqref{pdnimpedanceconstraint} as described has the decision variable formulation:

\begin{equation}\label{pdnbasicdecision}
\begin{split}
    \text{select} \quad N_{ij} \quad \forall i\text{,}j \quad \text{where} \quad i \neq j
\end{split}    
\end{equation}

However, we can trivially expand our selection space to also include:

\begin{equation}\label{pdnadvdecision}
\begin{split}
    \text{select} \quad Y_{ij} \quad \forall i\text{,}j \quad \text{where} \quad i \neq j
\end{split}    
\end{equation}

In this more complicated formulation, the optimization model selects admittances between capacitor placement areas. Beyond the selection space expansion, we must make a corresponding constraint and objective function formulation. 

In our PDN design, the admittance between ports will be a function of the copper geometry between the ports, including proximity to return planes. Heuristically, we may simplify the issue to:

\begin{equation}\label{pdnbasicdecision}
\begin{split}
    Y_{ij} = f(D_{ij}, W_{ij})
\end{split}    
\end{equation}

where $D_{ij}$ is the distance between ports (fixed with respect to the model), and $W_{ij}$ is the copper width of the run of copper between ports. The implementation of the function $f$ is left for further work, but should account for factors of the design such as the PCB stackup. Especially at higher frequencies the admittance will depend on more complicated geometric interactions that cannot be modeled in such a simple optimization model. 

In this formulation, we depend on the designer to assign preference $L_{ij}$ weights to each $W_{ij}$ relative to $K$ in equation \eqref{pdnobjfn} in order to scalarize it, making the new objective function:

\begin{equation}\label{pdnobjfn}
    \sum_{j}^{J}\sum_{i}^{I}\left (  (K_ja_i+b_i)N_{ij} + L_{ij}W_{ij} \right )
\end{equation}

\section{Quantitative part demand exercise}

Electronic designers often need to evaluate MLCC suppliers' part offerings for potential to use in their designs. By scaling the methodology presented in this paper to an entire design, we can precisely determine our quantity demanded for a part at a given price, or our \textit{capacitor demand curve} for an entire design or multiple designs.

Suppose we have $P$ circuit applications, each with an efficient solution vector $N_p$ for each circuit. Our demand curve for a part $i$ can be written as:

\begin{equation}\label{designsum}
    Q(C_i)=\sum_{p}^{P}N_{pi}
\end{equation}

where $N_{pi}$ is the number of capacitors of part $i$ used in application $p$ when its cost is $C_i$. The shape of $Q(C_i)$ must be weakly decreasing. Since the demand curve is a series of sums of optimization results, we can understand it as encoding important information about the opportunity cost of this part relative to others in our database.

We can use the generated demand curve in several ways:

\begin{itemize}
  \item By finding the intersection with our supply curve (e.g. volume discount), we can determine the optimal quantity of the part used.
  \item Knowing the x-intercept of the demand curve, we can quickly eliminate certain parts based on their suppliers' quoted price.
  \item By calculating the area under the demand curve and above the price, we can estimate the whole-device cost savings by introducing a new part. This savings accurately accounts for the opportunity cost of the part swaps.
\end{itemize}

\begin{figure}[htbp]
\centerline{\includegraphics[width=8cm]{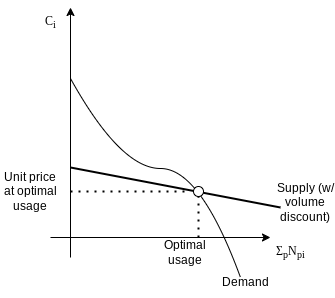}}
\caption{Supply and demand model for a single part $i$ within an electronic design.}
\label{figdemand}
\end{figure}

Figure \ref{figdemand} illustrates the supply and demand model for a single part i within an electronic design. It shows how the generated demand curve for a part can be used to find the intersection with the supply curve the firm faces (which may include volume discounts) to determine the optimal quantity of the part to be used. The graph also depicts the unit price at this optimal usage level.

\section{Discussion}

We have presented a spectrum of optimization frameworks, from simple linear $C_{\text{eff}}$ optimization to non-linear triple-objective models with geometric placement area considerations. For each framework presented, we have commented on the usefulness and accuracy of the selection methods, with the simplest models being the most useful in the current state of the art, and the more complicated models needing more future investment to become useful.

We believe there is immense opportunity to leverage these models to economize and efficiently pack capacitors in electronic designs. The quality of the model's output is not dependant on the human designer's selection intuition, but rather on the accuracy of a part library database and the appropriate electrical constraints.

In future works, we hope to examine the appropriateness of the heuristics \eqref{parallelapprox} and \eqref{pdnbasicdecision} in a wide range of applications. We also hope to integrate capacitor selection models into larger auto-designing workflows.

\end{document}